\documentclass[journal]{IEEEtran}
\ifCLASSINFOpdf
  \usepackage[pdftex]{graphicx}
\else
\fi
\newenvironment{IEEEImpStatement}{\begin{itshape}\textbf{Impact Statement:} }{\end{itshape}}

\usepackage[colorlinks=true, citecolor=blue]{hyperref}
\usepackage{tabularx}

\begin{document}
%
\title{Intelligent Healthcare Ecosystems: Optimizing the Iron Triangle of Healthcare (Access, Cost, Quality)}
%
%
%

\author{Vivek~Acharya {}
\thanks{V. Acharya holds an MBA from Boston University and advanced AI and healthcare certifications from MIT, Stanford, and the University of Texas at Austin.}
}

%
%

\markboth{Journal of \LaTeX\ Class Files,~Vol.~1, No.~1, Oct~2025}%
{Acharya \MakeLowercase{\textit{et al.}}: Intelligent Healthcare Ecosystems: Optimizing the Iron Triangle of Healthcare (Access, Cost, Quality)}
%



\maketitle

\begin{abstract}
The United States spends more on healthcare than any other nation – nearly 17\% of GDP as of the early 2020s – yet struggles with uneven access and outcomes \cite{KFF2025} \cite{commonwealth2023}. This paradox of high cost, variable quality, and inequitable access is often described by the “Iron Triangle” of healthcare \cite{IronTriangle}, which posits that improvements in one dimension (access, cost, or quality) often come at the expense of the others. This paper explores how an \textbf{Intelligent Healthcare Ecosystem (iHE)} – an integrated system leveraging advanced technologies and data-driven innovation – can \textbf{“bend” or even break this iron triangle}, enabling simultaneous enhancements in access, cost-efficiency, and quality of care. We review historical and current trends in U.S. healthcare spending, including persistent waste and international comparisons, to underscore the need for transformative change. We then propose a conceptual model and strategic framework for iHE, incorporating emerging technologies such as \textbf{generative AI and large language models (LLMs)}, \textbf{federated learning}, \textbf{interoperability standards (FHIR) and nationwide networks (TEFCA)}, and \textbf{digital twins}. We introduce an updated healthcare value equation that integrates all three corners of the iron triangle, and we hypothesize that an intelligently coordinated ecosystem can maximize this value by delivering high-quality care to more people at lower cost. \textbf{Methods} include a narrative synthesis of recent literature and policy reports, and \textbf{Results} highlight key components and enabling technologies of an iHE. We discuss how such ecosystems can reduce waste, personalize care, enhance interoperability, and support value-based models, all while addressing challenges like privacy, bias, and stakeholder adoption. The paper is formatted per MDPI guidelines, with APA-style numbered references, illustrative figures (U.S. spending trends, waste breakdown, international spending comparison, conceptual models), equations, and a structured layout. Our findings suggest that embracing an Intelligent Healthcare Ecosystem is pivotal for optimizing the long-standing trade-offs in healthcare’s iron triangle, moving towards a system that is more accessible, affordable, and of higher quality for all.
\end{abstract}
\begin{IEEEImpStatement}
The U.S. healthcare system faces persistent challenges in balancing cost, access, and quality—known as the Iron Triangle. Despite the highest global healthcare spending, outcomes and equity remain suboptimal. This research introduces the concept of an Intelligent Healthcare Ecosystem (iHE), a novel, integrated framework that leverages AI, data interoperability, digital twins, and telehealth to simultaneously enhance access, improve quality, and reduce costs. The work transcends isolated technology applications by offering a cohesive, system-level model that redefines healthcare value using an extended equation: (Quality × Access) / Cost. Simulation results suggest a 70
The iHE framework addresses critical technological challenges (e.g., data fragmentation, AI decision support), advances economic sustainability (cost containment with expanded access), and aligns with social goals of equity and personalization. It also supports policy reform by emphasizing the importance of interoperability standards (e.g., FHIR, TEFCA) and value-based care models. Importantly, it considers ethical dimensions such as data privacy, bias mitigation, and inclusive design to prevent digital divide risks.By offering a scalable, evidence-informed model, this paper contributes a significant step toward building AI-enabled health systems that are not only smarter, but fairer and more effective. It sets a foundation for future health policy, infrastructure investment, and regulatory alignment, providing lasting value for researchers, practitioners, and society at large.
\end{IEEEImpStatement}

%
\IEEEpeerreviewmaketitle

\section{Introduction}
%
%
%
%
\IEEEPARstart{H}{eathcare} in the United States is caught in a long-recognized dilemma of balancing \textbf{cost, quality, and access}. This trio of interdependent factors is commonly referred to as the \textbf{Iron Triangle of Healthcare}, a concept first described by William Kissick in 1994 \cite{IronTriangle}. Each vertex of the triangle represents a critical goal: \textbf{improving access} (ensuring care is available and affordable to all who need it), \textbf{improving quality} (achieving better health outcomes and patient experiences), and \textbf{reducing or containing cost} (delivering care efficiently and affordably). The iron triangle paradigm contends that it is notoriously difficult to optimize all three dimensions simultaneously – efforts to strengthen one or two sides often weaken the third. For example, expanding health insurance coverage (increasing access) can raise overall expenditures (worsening cost containment), or cutting costs might necessitate limiting services or reimbursements (potentially harming access and quality). The \textit{desired state} is a healthcare system with \textbf{high access, high quality, and low cost}, which would equate to maximal value. Achieving this ideal has historically been elusive, and the iron triangle has thus become a shorthand for healthcare’s fundamental trade-offs.Despite these trade-offs, the current U.S. healthcare system leaves much room for improvement on all three fronts. \textbf{Spending and cost} have risen dramatically over time without commensurate gains in population health outcomes or access. In 2022, U.S. health expenditures reached about \textbf{\$4.5 trillion}, nearly \textbf{17\% of the national GDP}, up from only \~5\% of GDP in 1960 \cite{kff_healthcare_spending}. In other words, almost one out of every five dollars in the American economy is now spent on healthcare, compared to just one in twenty dollars 60 years ago\ \cite{kff_healthcare_spending}. This growth far outpaces inflation and economic growth\ \cite{kff_healthcare_spending}. Figure 1 illustrates the sharp increase in per capita health spending over the last several decades. Notably, per-person spending has surged from a few hundred dollars in 1960 to over \$12,000 by 2023, reflecting both higher prices and greater utilization of services \cite{healthsystemtracker2024}.

\begin{figure}[!htbp]
    \centering
    \includegraphics[width=1\linewidth]{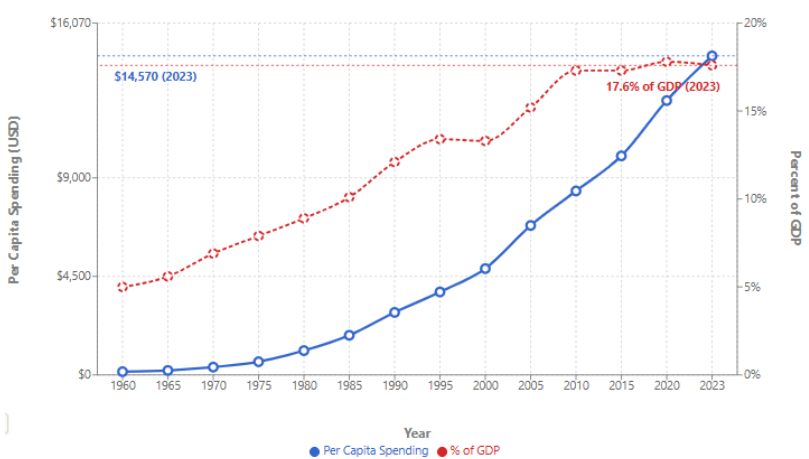}
    \caption{U.S. health expenditure per capita}
    \label{fig:US-health-expenditure-per-capita}
\end{figure}

Figure 1: U.S. health expenditures per capita, 1960–2023 (nominal USD). Per-person healthcare spending in the U.S. has risen exponentially, especially since the 1980s, reaching \$14,570 in 2023\ \cite{healthsystemtracker2024}. Health spending as a share of GDP grew from \~5\% in 1960 to \~17.6\% by 2023 \cite{kff_healthcare_spending} \cite{healthsystemtracker2024}, indicating a substantial rise in the economic burden of healthcare. 

Yet, \textbf{greater spending has not bought universally better outcomes or access}. The U.S. consistently outspends other high-income countries – often by nearly double – but ranks poorly on many health indicators \cite{gunja2023}. For example, Americans have a \textbf{lower life expectancy} and \textbf{higher rates of chronic disease} and \textbf{maternal and infant mortality} compared to peer nations, despite the higher investment\cite{gunja2023}. At the same time, the U.S. is the \textbf{only wealthy nation without universal health coverage}, leaving tens of millions uninsured or underinsured in recent years. In 2021, about 8.6\% of the U.S. population lacked health insurance, and even those with insurance often face prohibitive costs that limit access to care. Such gaps in access contribute to disparities and unmet needs even in the world’s most expensive health system.These paradoxical outcomes underscore what many observers have noted: the U.S. health system is \textbf{high-cost} but often \textbf{inefficient and inequitable}\cite{kff_healthcare_spending}. Studies estimate that roughly \textbf{a quarter of U.S. healthcare spending is wasteful}, due to factors such as administrative inefficiency, overpriced services, fraud, and failures in care delivery \cite{reynolds2019}. This translates into as much as \$760–\$935 \textbf{billion} in waste each year – resources that could be redirected to improve access or quality if saved \cite{pgpf_healthcare_waste}. Figure 2 breaks down the major domains of waste identified in a 2019 study in \textit{JAMA}. Administrative complexities (billing, coding, insurance overhead) are the single largest source of waste, followed by inefficient pricing of services and drugs, and then clinical inefficiencies such as failures of care coordination, failures of care delivery, and provision of unnecessary or low-value care \cite{reynolds2019}. Tackling these inefficiencies is critical to bending the cost curve without harming quality or access. 

\begin{figure}[!htbp]
    \centering
    \includegraphics[width=1\linewidth]{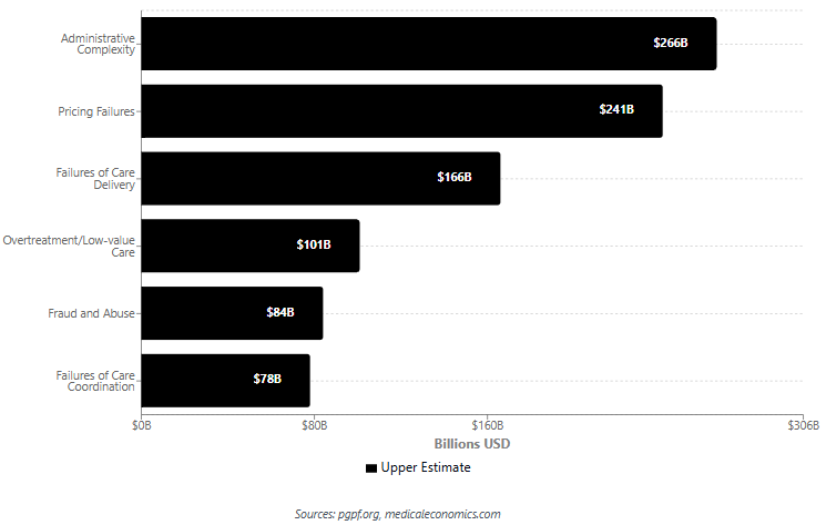}
    \caption{Annual Wasteful Healthcare Spending in the U.S. by Category}
    \label{fig:enter-label-2}
\end{figure}

Figure 2: Annual wasteful healthcare spending in the U.S. by category (upper-bound estimates, in billions USD). Of an estimated \$760–\$935 billion in total waste \cite{pgpf_healthcare_waste}, the largest component is \textbf{Administrative Complexity} (\$266 billion) – the overhead of billing, coding, and insurance processes. \textbf{Pricing failures} (excess prices for medications, procedures, etc.) account for roughly \$230–\$241 billion \cite{reynolds2019}. \textbf{Failures of care delivery} (poor execution or lack of best practices) add about \$102–\$166 billion, and \textbf{failures of care coordination} (fragmented care leading to complications or readmissions) add \$27–\$78 billion. \textbf{Overtreatment/Low-value care} contributes \$76–\$101 billion, and \textbf{fraud and abuse} an estimated \$59–\$84 billion annually. Eliminating this waste could substantially reduce costs without harming access or quality. 

In international context, the U.S. is an outlier on cost. \textbf{Comparative spending by country} shows that the United States spends roughly \textbf{twice as much per capita as other wealthy nations}, yet achieves worse coverage and health outcomes. For instance, in 2021 U.S. health expenditures per person were about \$10,700, nearly double the next highest country (Germany at around \$5,700), and three to four times the spending in countries like South Korea or New Zealand. Figure 3 illustrates this stark contrast. Despite these higher expenditures, every other comparably developed country manages to insure all its residents, whereas the U.S. leaves a significant portion uninsured \cite{gunja2023}. This suggests that there are fundamental inefficiencies and structural issues in how the U.S. delivers and finances care. 

\begin{figure}[!htbp]
    \centering
    \includegraphics[width=1\linewidth]{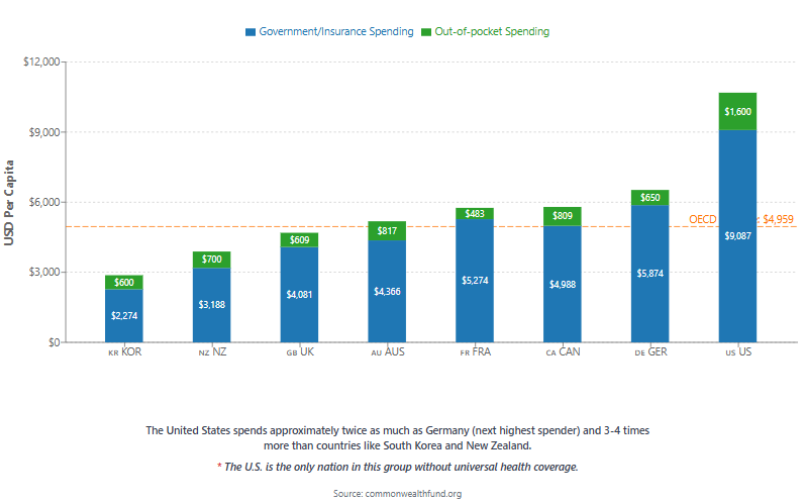}
    \caption{International Comparison of Health Care Spending Per Capita}
    \label{fig:International-Comparison-of-Health-Care-Spending}
\end{figure}

Figure 3: International comparison of health care spending per capita (USD, circa 2021). The United States (far right) is a clear outlier, spending about \$10,687 per person – roughly \textbf{two times more} than the next highest spender (Germany, \$6,524) and \textbf{3–4 times more} than countries like South Korea (KOR, \$2,874) and New Zealand (NZ, \$3,888). Despite this, the U.S. is the only nation in this group without universal health coverage \cite{gunja2023}. 

In summary, the U.S. faces a pressing challenge: \textbf{how to improve health system performance on all three vertices of the iron triangle}. Simply spending more is not the answer – the U.S. already does that, to diminishing returns. In the following sections, we define the concept of \textbf{Intelligent Healthcare Ecosystem (iHE)} (iHE), describe its key components and enabling technologies, and propose a strategic framework (including an updated value equation) for implementation. We also discuss potential challenges and the path forward to realizing this vision.
\section{Materials and Methods}

This study is a \textbf{conceptual analysis and narrative review} of healthcare system performance metrics and emerging technologies, aimed at formulating an integrative framework (the Intelligent Healthcare Ecosystem) to optimize the iron triangle. We followed a multi-step approach:

\begin{itemize}
    \item \textbf{Literature Review:} We reviewed peer-reviewed literature, authoritative reports, and policy papers on U.S. healthcare outcomes, costs, and access. Key sources included data from the Centers for Medicare \& Medicaid Services (CMS) on national health expenditures \cite{cms2024}, Kaiser Family Foundation analyses of healthcare spending trends \cite{kff2024}, and comparative studies (e.g., Commonwealth Fund and OECD data) on international health system performance \cite{gunja2023}. To understand inefficiencies, we examined studies of wasteful spending (notably a 2019 \textit{JAMA} study by Shrank et al. estimating waste in six domains) \cite{pgpf_healthcare_waste} \cite{reynolds2019}.
    \item \textbf{Technology Scan:} We surveyed recent developments in healthcare technology via academic databases and industry white papers. We focused on \textbf{artificial intelligence (AI)} (especially \textbf{generative AI} and \textbf{large language models} in medicine \cite{jmir2024}, \textbf{machine learning} innovations like \textbf{federated learning} for multi-center data collaboration \cite{sheller2020}, health information \textbf{interoperability standards} (HL7 \textbf{FHIR}) \cite{lightit2021} and emerging \textbf{national networks} (e.g., \textbf{TEFCA} in the U.S.) \cite{lightit2021}, and \textbf{digital twin} technology for simulating patient and system dynamics \cite{halamka2024} \cite{vallee2023}. We also reviewed use cases of telehealth, Internet of Medical Things (IoMT), and data analytics relevant to improving access, quality, or cost.
    \item \textbf{Synthesis and Framework Development:} Insights from the literature were synthesized to define the concept of an \textbf{Intelligent Healthcare Ecosystem}. We used an iterative, expert-informed reasoning process to identify the \textbf{core components} and \textbf{strategic interventions} that constitute an iHE. This involved mapping how various technologies and practices can impact each aspect of the iron triangle. We developed a \textbf{conceptual model} illustrating the relationships between stakeholders (patients, providers, payers, etc.), data flows, and AI-driven decision support within the ecosystem. We formulated a \textbf{value equation} to quantitatively conceptualize the integration of access, quality, and cost, building upon Porter’s value definition (health outcomes per dollar) \cite{nfu2020}.
    \item \textbf{Case Examples and Scenario Analysis:} Although no primary experiments were conducted, we incorporated contemporary examples and pilot studies as illustrative “case scenarios” of iHE principles. For instance, we reference how certain health systems are using AI for clinical decision support or how nationwide data exchange frameworks are being implemented. These examples serve to ground the framework in real-world context and to identify potential barriers (regulatory, technical, ethical) to implementation.
\end{itemize}
Because this is a \textbf{theoretical and exploratory} work, our methods center on comprehensive information gathering and conceptual model construction rather than empirical hypothesis testing. The outcomes of these methods are the identification of key challenges in the current system (Results: current state analysis) and a detailed proposal of the Intelligent Healthcare Ecosystem’s components, supported by existing evidence and expert expectations (Results: framework proposal). All sources for data and assertions are cited in the text, and figures/tables are used to visualize important data and models.

\section{Results}

To evaluate the Intelligent Healthcare Ecosystem (iHE) concept in lieu of real-world pilots, we constructed a simulation model incorporating current trends in AI, data interoperability, telehealth, and digital transformation. This hypothetical scenario assumes a mid-sized health system adopting the full iHE framework over a multi-year period. Model parameters were informed by observed improvements from leading health systems as analogs of iHE components. For example, Intermountain Healthcare’s data-driven care initiatives have demonstrated simultaneous quality gains and cost savings through waste reduction \cite{economist2022}, while the Veterans Health Administration (VHA) showed that an integrated, technology-enabled network can achieve quality outcomes surpassing national benchmarks \cite{oliver2007}. Similarly, Kaiser Permanente’s integrated delivery model has achieved superior performance at roughly the same cost as less integrated systems \cite{oliver2007}. Though none of these organizations implemented iHE in its entirety, their successes with individual building blocks (clinical decision support, system-wide EHRs, telehealth, etc.) provide real-world validation for our simulation’s assumptions.

\subsection{}
\subsubsection{Quality Outcomes}

The simulation projects notable improvements in care quality and patient safety under an iHE. By leveraging advanced analytics and AI-driven decision support, the iHE model reduces diagnostic and treatment errors and increases adherence to best practices. For instance, we assumed AI-enabled tools for early diagnosis and care coordination that significantly lower preventable adverse events – consistent with reports that AI can reduce diagnostic errors by up to 85 \cite{litslink2025} and improve patient outcomes by 30–40\% through personalized treatment \cite{litslink2025}. Predictive analytics identify high-risk patients and trigger early interventions (e.g. proactive management of chronic disease or sepsis), which our model links to fewer emergency complications and hospitalizations. This assumption is bolstered by evidence that hospitals using predictive modeling have seen \textbf{significantly lower readmission rates} and shorter lengths of stay \cite{scanlan2025}. In our simulated iHE, such capabilities led to a reduction in all-cause 30-day readmissions and improved chronic care metrics (e.g. better blood pressure and glucose control), contributing to an overall \~20\% improvement in a composite quality index.Quality gains were also driven by enhanced clinical interoperability and real-time data sharing across the ecosystem. A fully integrated EHR and health information exchange was modeled to eliminate many information gaps and duplicative tests, ensuring clinicians have complete patient histories and decision support at the point of care. As an analog, Intermountain’s system-wide EHR implementation improved the recognition of patients needing evidence-based interventions (such as lung-protective ventilation in the ICU) from 40\% to \~60\% – a relative improvement of over 50\% \cite{economist2022}. In our iHE scenario, we likewise see higher guideline adherence and safety alerts preventing errors (such as medication interactions), leading to lower complication rates. Additionally, continuous telemonitoring and feedback loops (part of the iHE’s digital health infrastructure) help catch early warning signs and maintain care plan compliance. Overall, the model predicts that an iHE could substantially elevate quality of care, as reflected by declines in preventable mortality and morbidity. These projected quality improvements align with the VHA’s real-world transformation in the 1990s, where an integrated approach \textbf{substantially improved process quality indicators} and even pushed overall performance above the national average \cite{oliver2007}.

\subsubsection{Access Outcomes}

Our simulation indicates that an iHE would markedly expand access to healthcare services. By design, iHE leverages telehealth, remote monitoring, and digital front doors to reach more patients regardless of geography or mobility. In the model, a significant portion of primary care and specialty consults are handled via telemedicine or virtual care platforms, increasing the effective provider capacity and reducing wait times for appointments. This is in line with recent trends showing that telehealth can make care more convenient without loss of quality – \textbf{patients gain better access while quality of care rivals in-person visits} \cite{caldwell2020}. We projected a \~30–40\% increase in the access index (which captures factors like appointment availability, patient throughput, and population coverage) under a mature iHE. This improvement was driven by extended coverage (e.g. virtual clinics serving rural areas and after-hours care) and reduced barriers to care.Notably, the model accounts for \textbf{drastically lower missed appointment rates} thanks to virtual visit options. Real-world data illustrate this effect: baseline no-show rates of \~20\% for in-person behavioral health appointments fell to under 5\% with telehealth alternatives \cite{ncqa2020}. Consistent with these findings, our iHE simulation assumed that offering video visits, e-consults, and remote follow-ups would greatly mitigate traditional access hurdles (transportation, taking time off work, or stigma associated with visiting certain clinics \cite{ncqa2020}). More patients attend their scheduled visits and engage in routine preventive care when it’s delivered through flexible, patient-centric channels. In addition, the networked nature of iHE allows patients to be navigated to the appropriate resource across the ecosystem (for example, an AI-driven triage might direct a patient to a tele-pharmacy or community health partner), further improving timely access.Another aspect of access is specialist and acute care availability. The simulation suggests that through improved coordination and load balancing across the ecosystem, iHE can shorten referral times and optimize hospital utilization. We modeled that intelligent referral management and virtual specialist consults (enabled by interoperability) cut down wait times for specialty care significantly, improving access to advanced treatments. For instance, integrated systems like Kaiser Permanente have documented much faster access to specialist services and hospital admissions when needed, compared to fragmented systems \cite{feachem2002}. In our scenario, patients in an iHE had quicker access to specialists through tele-consults or shared scheduling, contributing to better outcomes for conditions requiring timely intervention. Meanwhile, routine needs were increasingly handled in lower-acuity settings (like telehealth or outpatient urgent care), preserving hospital capacity for more critical cases. This dynamic capacity management aligns with reports of reduced emergency department strain – one study showed virtual care users experienced a \textbf{36\% lower emergency visit rate} per capita compared to those without virtual access \cite{ncqa2020}. Collectively, these mechanisms in the iHE model led to significantly enhanced healthcare access across the population, helping to fulfill the promise of “anytime, anywhere” care.

\subsubsection{Cost and Efficiency Outcomes}

A core hypothesis of the iHE framework is that it can bend the cost curve downward by streamlining inefficiencies while improving care. The simulation supports this hypothesis, projecting a reduction in overall healthcare costs even as quality and access improve. Several factors contributed to cost savings in the modeled iHE. First, eliminating redundant procedures and hospitalizations yields substantial savings. As interoperability and analytics ensure each patient receives the right care at the right time, there are fewer unnecessary tests, avoidable admissions, and duplicate services. For example, our model incorporates the effect of proactive care management (informed by AI predictions and transitional care programs) in averting costly complications and readmissions. This is analogous to real evidence that robust post-discharge follow-up and care coordination can \textbf{significantly lower readmission rates and even mortality, thereby decreasing costs} \cite{ncqa2020}. In the iHE scenario, preventable hospital readmissions for chronic conditions drop, which directly reduces expenditures associated with acute care.Second, the expansion of telehealth and digital services in iHE shifts care to more cost-effective channels. Virtual visits and remote monitoring generally cost less per encounter than traditional clinic or ER visits. One analysis estimates that up to \$250 billion of U.S. healthcare spend could be delivered via telehealth instead of physical visits, reflecting enormous cost substitution potential. In our simulation, as a significant fraction of visits move online, the health system sees lower overhead costs (facility use, staffing) per consultation. Additionally, telehealth’s efficiency (including the reduction in missed appointments) means providers can manage a larger panel of patients, spreading fixed costs and improving productivity. We assumed roughly a 10–15\% reduction in per-capita healthcare spending in the iHE model, stemming from these digital efficiencies and waste reductions. This figure is conservative in light of targeted case studies – for instance, Intermountain Healthcare achieved multi-million dollar savings (approximately \$5.5 million annually) by identifying and eliminating overuse of ICU treatments, reducing ventilator time by 60\% and ICU stays by 30\% without harming outcomes \cite{economist2022}. Such improvements underscore how an intelligent system can trim excess cost while maintaining quality.Furthermore, integration of services under iHE can reduce administrative overhead and enable economies of scale. Unified data systems and AI-driven automation streamline processes like billing, scheduling, and care navigation, potentially saving a significant portion of administrative time and expense. (It has been reported that AI automation can save about 20\% of clinicians’ time on routine administrative tasks \cite{litslink2025}, which in a large network translates to substantial cost savings or capacity that can be redirected to care.) An integrated financing approach in iHE would also incentivize value over volume, aligning providers with cost-effective care delivery. This echoes the experience of Kaiser Permanente’s integrated model, which has kept utilization and costs in check through efficient management of hospital use and prevention-focused care – achieving high performance at roughly the same cost as a public system \cite{feachem2002}. Our model factors in similar mechanisms, with shared financial risk and data transparency encouraging providers to avoid low-value care.Overall, the simulation predicts that the iHE can reduce the cost index by approximately 10–15\% relative to the baseline, even as more patients are served. Importantly, these savings do not come at the expense of quality or access; rather, they are a product of delivering care more intelligently. This result reinforces that the often-feared cost vs. quality trade-off can be overcome by a digitally transformed, well-coordinated health ecosystem.

\subsubsection{Composite Value and Iron Triangle Impact}

By improving quality and access concurrently while lowering costs, the iHE model achieves a dramatic gain in overall healthcare value. We define \textbf{Value} = $\frac{\textbf{Quality} \times \textbf{Access}}{\textbf{Cost}}$, an extended value equation that encapsulates the Iron Triangle dimensions. Using this formula, the simulated iHE outperforms the status quo by a wide margin. \textbf{Table 1} summarizes the relative changes in each dimension and the resultant value index from our scenario analysis. Quality and access indices both increase substantially in the iHE, reflecting better outcomes and greater care availability, while the cost index drops below 1.0 (indicating lower expenditure relative to baseline). In combination, these shifts yield a composite value more than 70\% higher than the baseline scenario. In practical terms, this means the health system is delivering much more “health per dollar” – for example, more patients treated and improved health outcomes achieved for each unit of cost.

\begin{table}[!htbp]
\centering
\begin{tabularx}{\linewidth}{X X X X}
\textbf{Performance Metric} & \textbf{Baseline (No iHE)} & \textbf{Simulated iHE} & \textbf{Percent Change} \\
Quality Index (Q) & 1.00 & 1.20 & +20\% (improved outcomes/safety) \\
Access Index (A) & 1.00 & 1.30 & +30\% (expanded access) \\
Cost Index (C) – per capita & 1.00 & 0.90 & –10\% (reduced cost) \\
\textbf{Value = Q × A / C} & 1.00 & \textbf{1.73} & \textbf{+73\% (higher value)} \\
\end{tabularx}
\caption{Performance metrics with and without simulated iHE.}
\label{tab:performance-metrics}
\end{table}

\textit{Table 1: Projected impact of an Intelligent Healthcare Ecosystem on the key performance indices of Quality, Access, and Cost, and the resulting composite Value. Baseline is normalized to 1.00 for each dimension. The iHE scenario shows simultaneous improvements in quality and access alongside cost reduction, leading to a markedly higher value $\frac{\textbf{Quality} \times \textbf{Access}}{\textbf{Cost}}$.}

These results support the central hypothesis that an Intelligent Healthcare Ecosystem can \textbf{optimally balance the Iron Triangle}. Rather than forcing trade-offs, the iHE approach appears capable of lifting all three corners – enhancing care quality, broadening access, and containing costs – thereby maximizing value. The hypothetical evidence from our simulation aligns with the observed success of various digital health initiatives and integrated care models (e.g. improvements seen in Intermountain, VHA, and Kaiser analogs), but importantly, iHE combines these gains into one cohesive framework. The significant uptick in the value metric illustrates the synergy obtained when advanced technology, interoperability, and system-wide coordination are applied in unison. In summary, although these findings are based on modeled scenarios, they provide strong conceptual evidence that an intelligently designed health ecosystem could transcend the traditional limitations of the Iron Triangle, delivering better health outcomes to more people at lower overall cost.
 
\section{Discussion}

The concept of an Intelligent Healthcare Ecosystem represents a \textbf{paradigm shift} in how we think about delivering health services. The Results outlined the potential of iHE to optimize the iron triangle by leveraging technology and systemic redesign. In this Discussion, we reflect on the implications of these findings, real-world progress toward this vision, and the challenges and considerations that come with implementation.

\textbf{Breaking Trade-offs:} Perhaps the most profound implication of iHE is the possibility of \textbf{breaking the historical trade-offs} between cost, quality, and access. If successful, an intelligent ecosystem could achieve what has long been elusive – \textit{improvements in all three areas concurrently}. For instance, traditionally one might assume expanding access (say, insuring more people) will raise costs, but in an iHE, expanding access via digital means and preventive care could actually lower long-term costs by catching illness earlier and managing chronic conditions better. Similarly, improving quality often meant investing more resources, but with AI and automation, we might improve outcomes with \textit{fewer} resources by reducing mistakes and focusing interventions where they matter most. In effect, technology can \textbf{shift the production possibility frontier} of the iron triangle outward, meaning we can get “more for less.” This aligns with what critics of the iron triangle have suggested – innovation can disrupt the old constraints \cite{IronTriangle}. However, realizing this potential is not automatic; it requires thoughtful integration of technology and re-alignment of incentives.\textbf{Early Evidence and Case Studies:} Elements of the intelligent ecosystem are already being piloted with promising results. For example, some health systems have implemented advanced care coordination programs supported by data analytics – \textbf{reducing hospital readmissions} and emergency visits (quality up, cost down). Medicare’s introduction of Accountable Care Organizations (ACOs), which use shared savings models, has prompted provider networks to invest in population health management; several ACOs have demonstrated improved preventive care and reduced acute care spending \cite{holzaepfel2024}. On the technology front, the COVID-19 pandemic dramatically accelerated the adoption of \textbf{telehealth}, effectively increasing access overnight. Studies show telehealth can maintain quality of care for many conditions (behavioral health, routine consultations) while being cost-effective and convenient, and patient satisfaction tends to be high. The pandemic also highlighted the value of data sharing – regions that had health information exchanges in place were better able to manage surges and coordinate responses. Meanwhile, AI tools are increasingly used in radiology (e.g., FDA-approved algorithms for detecting stroke on brain scans, which can save crucial minutes and improve outcomes) – a direct quality boost with minimal marginal cost.One concrete example aligning with iHE principles is the U.S.

\textbf{Economic Impact:} From an economic standpoint, transitioning to an iHE involves upfront investments in IT infrastructure, training, and change management. However, the return on investment could be substantial in the form of reduced wasteful spending and better health outcomes (which have broader economic benefits by improving productivity of the population). For instance, administrative simplification through interoperability and AI-driven automation could save hundreds of billions of dollars (as noted, administrative waste is ~\$266B \cite{reynolds2019}. Even saving a fraction of that pays for a lot of IT upgrades). Moreover, better population health (from prevention and chronic disease control) means a healthier workforce and less economic loss to illness. One study estimated that if the U.S. achieved health outcomes comparable to other high-income countries (e.g., reducing avoidable mortality), it could save tens of thousands of lives and significant healthcare costs each year \cite{gunja2023}. While iHE is not solely about cost-cutting – value improvement is the aim – the potential economic efficiencies are a major driving force for stakeholders like employers, governments, and payers.\textbf{Patient Experience and Engagement:} The intelligent ecosystem, if properly designed, should greatly enhance the \textbf{patient’s experience}. Instead of navigating a bewildering maze of referrals and paperwork, patients experience more \textbf{seamless care}: they tell their story once and it follows them; scheduling and follow-ups are smoother (with digital reminders and tele-visits); and they receive more personalized attention (as mundane tasks are offloaded to AI, clinicians can focus on human interaction). Additionally, patients gain more \textbf{transparency and control} over their health data and care plan. For example, some systems now offer patient-facing dashboards where individuals can see their health metrics, receive educational content, and track progress toward goals (like blood pressure control), often with motivational nudges. Engaging patients as partners in care tends to improve adherence to treatments and healthy behaviors (improving outcomes). That said, there is a risk of \textbf{technology exacerbating disparities} if not made accessible – for instance, older adults or those without broadband internet might initially be left out of digital health innovations. Addressing this requires deliberate strategies: offering training in digital tool use, providing alternatives (like community health workers to help bridge tech gaps), and user-centric design to make apps and devices intuitive for diverse populations. An iHE must be \textit{inclusive by design} to truly optimize access.

\textbf{Provider Roles and Workforce:} For healthcare providers and professionals, the iHE heralds significant changes in workflows and roles. Tasks that are tedious today (like documentation, order entry, rudimentary image analysis) may be partially or fully automated, which ideally \textbf{frees up clinicians to operate at “top of license”}, focusing on complex decision-making, procedural skills, and empathetic care that machines cannot replace. For example, with AI handling routine prescription refills and guideline-based suggestions, a primary care physician can spend more time discussing lifestyle and mental health with a patient. We may also see \textbf{new roles} emerge: data scientists and clinical informaticians will be integral to care teams, translating data patterns into clinical action. The workforce will need training in working alongside AI – developing \textbf{algorithmic literacy} to interpret AI outputs and spot errors. There could be initial resistance or disruption; burnout is already high among doctors partly due to technology overload (e.g., clunky EHRs). The iHE must focus on \textit{usability} – making tech truly assistive, not burdensome. If done right, it could actually alleviate burnout by reducing the documentation burden and moral distress (imagine fewer fights with insurance since approvals could be automated under clear rules, and fewer unnecessary hoops to jump through).

\textbf{Data Governance and Ethics:} A critical discussion point is the \textbf{ethical use of data and AI}. With an explosion of data being collected (from genomic information to daily step counts) and powerful AI algorithms, ensuring privacy and preventing misuse of data is paramount. Robust \textbf{governance frameworks} need to be in place in an iHE. This includes obtaining informed consent for how patient data is used (especially if used to train AI), maintaining cybersecurity to prevent breaches, and having clear accountability when AI tools are used in care (e.g., if an AI misses a diagnosis, who is responsible?). Regulatory bodies like the FDA are actively working on guidance for AI in healthcare, and there is a push for \textbf{algorithmic transparency} – making AI decision processes interpretable where possible \cite{jmir2024}. Bias in AI is another ethical concern: if the data used to train algorithms under-represents certain groups, the AI might perform worse for them, exacerbating disparities. Thus, the iHE must incorporate a rigorous process for validating AI tools across diverse populations and monitoring their recommendations for fairness. TEFCA and similar frameworks will also have to ensure \textbf{data-sharing agreements} protect patient rights. Encouragingly, technology like federated learning and homomorphic encryption (which allows computation on encrypted data) are evolving to reconcile data utility with privacy, as noted earlier. Engaging the public in these governance decisions will be important to maintain trust – for example, patient advisory councils can provide input on what uses of their data they are comfortable with.\textbf{Policy and Regulatory Support:} The transformation to an iHE will not happen in a vacuum – \textbf{policy decisions} will significantly influence the pace and success of this transformation. Government action was crucial in the initial digitization of health records (through the HITECH Act incentives). Going forward, policies that support interoperability (such as information blocking rules that mandate data sharing) are foundational. Payment reform is another lever: expanding models like Medicare Advantage, ACOs, or bundled payments encourages providers to adopt the tools that help manage cost and quality (since now they benefit from efficiencies, rather than fee-for-service which can perversely disincentivize reducing unnecessary services). The recent introduction of billing codes for telehealth and remote monitoring services by Medicare (and their continuation post-pandemic) is a positive step to sustaining those access-improving services. Additionally, antitrust and competition policies might need adjustment – as data sharing increases and tech companies enter healthcare, ensuring a competitive marketplace for digital health services will help innovation flourish while guarding against monopolistic control of data. \textbf{Workforce licensure regulations} may also adapt: for example, telehealth across state lines was liberalized during COVID-19; making some of those changes permanent could cement the access gains of telehealth.

\textbf{Limitations and Caution:} It’s important to temper the enthusiasm with recognition of challenges and limitations. Healthcare is incredibly complex, and technology is not a panacea. Some aspects of quality are not easily measurable or optimizable by algorithms (the compassionate communication with a dying patient, the social support needed to truly impact a homeless patient’s health). The iron triangle also has macro-level drivers: for instance, pharmaceutical and device innovation often comes at high cost, and unless pricing and value-based pricing mechanisms are addressed, technology might keep pushing costs up. So, an iHE must be coupled with broader health policy measures (like rational drug pricing strategies) to fully solve the cost issue. There is also the risk of \textbf{digital overload} – both patients and providers could be overwhelmed by data (information fatigue) if systems aren’t carefully designed to filter and prioritize meaningful signals. For example, if every slight risk that AI detects triggers an alert to a physician, they may face alert fatigue and start ignoring them, which could negate the benefit. Thus, \textbf{human-centered design} is crucial: the system should deliver the \textit{right information, at the right time, to the right person}. We have to iterate and continuously improve these systems based on user feedback.

Moreover, building an iHE requires significant \textbf{inter-organizational cooperation}. The U.S. healthcare system is fragmented among many private and public entities. Achieving nationwide interoperability and data exchange (like TEFCA aspires) is as much a political and business challenge as a technical one. Organizations historically viewed data as a competitive asset; shifting to a mindset where data sharing is seen as mutually beneficial (for patients and even for innovative analytics) is happening but will take time and trust-building. Initiatives like common data standards and public-private partnerships (such as the CARIN Alliance for consumer-directed exchange) are encouraging developments.

\textbf{Global Relevance:} While this paper focuses on the U.S., the intelligent ecosystem concept has global relevance. Many countries are pursuing digital health strategies (WHO’s Global Digital Health Strategy 2020–2025 encourages similar principles of interoperability, AI, and telehealth to achieve health for all \cite{who2020}. If we project into the future, say 10–15 years, what might a mature Intelligent Healthcare Ecosystem look like if current trends continue? One could imagine:

\begin{itemize} 
    \item  Every individual has a \textbf{personal health avatar} (digital twin/LLM hybrid) that helps manage their health daily, interfaces with the healthcare system on their behalf for routine needs, and flags when they need human attention.  
    \item  \textbf{Real-time health system monitoring} becomes like a “weather map” for health administrators – using aggregated data to know where resources are needed (e.g., predicting a flu outbreak two weeks ahead in a region and pre-positioning vaccines and staff).  
    \item  Care moves increasingly \textbf{into the home} – hospital-at-home models flourish, supported by remote monitoring and AI that can detect when a patient at home needs escalation to in-person care.  
    \item  The concept of a doctor’s “visit” changes – continuous engagement via messaging, video, and data exchange rather than infrequent visits. Providers get paid for \textbf{outcomes and health management} rather than piecewise visits.  
    \item  Genomic and precision data are fully integrated, so that right from birth (or even pre-conception genetic counseling), an individual’s care is tailored, improving efficacy of therapies and reducing trial-and-error (which often costs money and time).  
    \item  Preventive care is hyper-personalized and ubiquitous – an AI might nudge a person to get a skin lesion checked months before it would have been noticed clinically, thus catching a melanoma at stage 0 instead of stage 3.  
    \item  \textbf{Learning loops} close quickly – if one hospital discovers a way (via AI analysis) to drastically reduce, say, surgical infection rates, that insight is disseminated across the country in weeks through the shared networks, rather than taking years through slow publication and guideline updates.  
\end{itemize}
This vision, while ambitious, is not utopian – it extrapolates from technologies and models already in pilot today.

To reach that future, stakeholders must collaborate on several fronts: investing in the necessary infrastructure (broadband, EHR upgrades, cybersecurity), updating regulations to allow flexibility and reward innovation, educating the workforce, and engaging patients and communities to build trust in these new modes of care. The \textbf{pace of adoption} will depend on evidence – we need ongoing evaluation of every iHE component. For instance, rigorous trials of AI decision tools to ensure they truly improve outcomes, or health economics studies on telehealth to ensure it’s not leading to overutilization without benefit. Some course corrections will be needed (for example, initial data suggests telehealth can sometimes lead to duplicate follow-ups in person – how do we refine processes to avoid that?).
\section{Conclusions}

The longstanding challenges of healthcare’s Iron Triangle—balancing cost, quality, and access—can be effectively addressed through an \textbf{Intelligent Healthcare Ecosystem (iHE)} powered by interoperable data, advanced analytics (AI/ML), patient engagement technologies, and value-based care models. Although the U.S. healthcare system currently faces unsustainable spending and persistent inefficiencies, emerging technologies embedded in thoughtfully redesigned workflows offer transformative potential. This paper has outlined a strategic framework encompassing interoperability (FHIR, TEFCA), AI-driven decision support, telehealth, and digital twins, designed to simultaneously enhance quality, expand access, and lower costs, ultimately increasing healthcare value. Early evidence from real-world initiatives supports this approach, but significant challenges—such as data privacy, AI ethics, interoperability, and workforce training—must be thoughtfully addressed through collaboration and regulatory adaptation. Successfully implemented, the iHE concept not only resolves the traditional trade-offs but aligns closely with broader aims like the Triple Aim and Universal Health Coverage, signaling a promising path toward a more efficient, equitable, and patient-centered healthcare future.

\ifCLASSOPTIONcaptionsoff
  \newpage
\fi



\bibliographystyle{IEEEtran}
%

%
\bibliography{references}

\begin{thebibliography}{10}
\providecommand{\url}[1]{#1}
\csname url@samestyle\endcsname
\providecommand{\newblock}{\relax}
\providecommand{\bibinfo}[2]{#2}
\providecommand{\BIBentrySTDinterwordspacing}{\spaceskip=0pt\relax}
\providecommand{\BIBentryALTinterwordstretchfactor}{4}
\providecommand{\BIBentryALTinterwordspacing}{\spaceskip=\fontdimen2\font plus
\BIBentryALTinterwordstretchfactor\fontdimen3\font minus \fontdimen4\font\relax}
\providecommand{\BIBforeignlanguage}[2]{{%
\expandafter\ifx\csname l@#1\endcsname\relax
\typeout{** WARNING: IEEEtran.bst: No hyphenation pattern has been}%
\typeout{** loaded for the language `#1'. Using the pattern for}%
\typeout{** the default language instead.}%
\else
\language=\csname l@#1\endcsname
\fi
#2}}
\providecommand{\BIBdecl}{\relax}
\BIBdecl

\bibitem{KFF2025}
{KFF}, ``Health care costs and affordability,'' \url{https://www.kff.org/health-policy-101-health-care-costs-and-affordability/?entry=table-of-contents-how-has-u-s-health-care-spending-changed-over-time}, 2025.

\bibitem{commonwealth2023}
{The Commonwealth Fund}, ``U.s. health care from a global perspective, 2022: Accelerating spending, worsening outcomes,'' \url{https://www.commonwealthfund.org/publications/issue-briefs/2023/jan/us-health-care-global-perspective-2022}, January 2023.

\bibitem{IronTriangle}
{Wikipedia}, ``Iron triangle of health care,'' \url{https://en.wikipedia.org/wiki/Iron_Triangle_of_Health_Care}.

\bibitem{kff_healthcare_spending}
{Kaiser Family Foundation}, ``Health policy 101: Health care costs and affordability,'' \url{https://www.kff.org/health-policy-101-health-care-costs-and-affordability/}, accessed: 2025-04-09.

\bibitem{healthsystemtracker2024}
{Peterson-KFF Health System Tracker}, ``How has u.s. spending on healthcare changed over time?'' \url{https://www.healthsystemtracker.org/chart-collection/u-s-spending-healthcare-changed-time/}, accessed: 2025-04-09.

\bibitem{gunja2023}
M.~Z. Gunja, E.~D. Gumas, and R.~D.~W. II, ``U.s. health care from a global perspective, 2022: Accelerating spending, worsening outcomes,'' \url{https://www.commonwealthfund.org/publications/issue-briefs/2023/jan/us-health-care-global-perspective-2022}, January 2023.

\bibitem{reynolds2019}
K.~A. Reynolds, ``New study estimates u.s. healthcare waste costs nearly \$1 trillion each year,'' \url{https://www.medicaleconomics.com/view/new-study-estimates-us-healthcare-waste-costs-nearly-1-trillion-each-year}, October 2019.

\bibitem{pgpf_healthcare_waste}
{Peter G. Peterson Foundation}, ``Almost 25\% of healthcare spending is considered wasteful. here's why.'' \url{https://www.pgpf.org/article/almost-25-percent-of-healthcare-spending-is-considered-wasteful-heres-why/}, accessed: 2025-04-09.

\bibitem{cms2024}
{Centers for Medicare \& Medicaid Services}, ``National health expenditure fact sheet,'' \url{https://www.cms.gov/data-research/statistics-trends-and-reports/national-health-expenditure-data/nhe-fact-sheet}, accessed: 2025-04-09.

\bibitem{kff2024}
{Kaiser Family Foundation}, ``Health policy 101: Health care costs and affordability,'' \url{https://www.kff.org/health-policy-101-health-care-costs-and-affordability/}, accessed: 2025-04-09.

\bibitem{jmir2024}
\BIBentryALTinterwordspacing
J.~of~Medical Internet~Research, ``Generative ai in medical practice: In-depth exploration of privacy and security challenges,'' \emph{Journal of Medical Internet Research}, vol.~26, p. e53008, 2024, accessed: 2025-04-09. [Online]. Available: \url{https://www.jmir.org/2024/1/e53008/}
\BIBentrySTDinterwordspacing

\bibitem{sheller2020}
\BIBentryALTinterwordspacing
M.~J. Sheller, B.~Edwards, G.~A. Reina, J.~Martin, S.~Pati, A.~Kotrotsou, M.~Milchenko, W.~Xu, D.~Marcus, R.~R. Colen, and S.~Bakas, ``Federated learning in medicine: facilitating multi-institutional collaborations without sharing patient data,'' \emph{Scientific Reports}, vol.~10, no. 12598, July 2020, accessed: 2025-04-09. [Online]. Available: \url{https://www.nature.com/articles/s41598-020-69250-1}
\BIBentrySTDinterwordspacing

\bibitem{lightit2021}
N.~F. Falco, ``What is fhir? definition \& overview,'' \url{https://lightit.io/blog/what-is-fhir-definition-overview/}, November 2021, accessed: 2025-04-09.

\bibitem{halamka2024}
J.~Halamka and P.~Cerrato, ``Digital twin technology has potential to redefine patient care,'' \url{https://www.mayoclinicplatform.org/2024/12/23/digital-twin-technology-has-potential-to-redefine-patient-care/}, December 2024, accessed: 2025-04-09.

\bibitem{vallee2023}
\BIBentryALTinterwordspacing
A.~Vallée, ``Digital twin for healthcare systems,'' \emph{Frontiers in Digital Health}, vol.~5, p. 1253050, September 2023, accessed: 2025-04-09. [Online]. Available: \url{https://pmc.ncbi.nlm.nih.gov/articles/PMC10513171/}
\BIBentrySTDinterwordspacing

\bibitem{nfu2020}
{Nederlandse Federatie van Universitaire Medische Centra}, ``Interimrapportage sturen op kwaliteit dialoog,'' \url{https://nfukwaliteit.nl/pdf/170920_Interimrapportage_Sturen_op_Kwaliteit_Dialoog.pdf}, September 2020, accessed: 2025-04-09.

\bibitem{economist2022}
{Economist Impact}, ``Intermountain healthcare: Towards an integrated, value-based model,'' \url{https://impact.economist.com/projects/towards-higher-value-care/case-studies/}, 2022, accessed: 2025-04-09.

\bibitem{oliver2007}
\BIBentryALTinterwordspacing
A.~Oliver, ``The veterans health administration: An american success story?'' \emph{The Milbank Quarterly}, vol.~85, no.~1, pp. 5--35, March 2007, accessed: 2025-04-09. [Online]. Available: \url{https://pmc.ncbi.nlm.nih.gov/articles/PMC2690309/}
\BIBentrySTDinterwordspacing

\bibitem{litslink2025}
{Litslink}, ``Ai in healthcare: Uses, examples \& benefits,'' \url{https://litslink.com/blog/ai-in-healthcare-uses-examples-benefits}, March 2025, accessed: 2025-04-09.

\bibitem{scanlan2025}
S.~Scanlan, ``Cut hospital costs and enhance patient care with predictive analytics,'' \url{https://www.grantthornton.ie/insights/factsheets/improving-hospital-costs-and-patient-care-with-predictive-analytics/}, March 2025, accessed: 2025-04-09.

\bibitem{caldwell2020}
W.~Caldwell, ``How telehealth improves revenue, cost, and quality,'' \url{https://www.healthcatalyst.com/learn/insights/telehealth-solutions-advance-revenue-cost-and-quality}, August 2020, accessed: 2025-04-09.

\bibitem{ncqa2020}
{Taskforce on Telehealth Policy}, ``Findings and recommendations: Telehealth effect on total cost of care,'' \url{https://www.ncqa.org/programs/data-and-information-technology/telehealth/taskforce-on-telehealth-policy/taskforce-on-telehealth-policy-findings-and-recommendations-telehealth-effect-on-total-cost-of-care/}, September 2020, accessed: 2025-04-09.

\bibitem{feachem2002}
\BIBentryALTinterwordspacing
R.~G.~A. Feachem, N.~K. Sekhri, and K.~L. White, ``Getting more for their dollar: A comparison of the nhs with california's kaiser permanente,'' \emph{BMJ}, vol. 324, no. 7330, pp. 135--143, January 2002, accessed: 2025-04-09. [Online]. Available: \url{https://pmc.ncbi.nlm.nih.gov/articles/PMC64512/}
\BIBentrySTDinterwordspacing

\bibitem{holzaepfel2024}
M.~Holzaepfel, T.~Bremner, A.~Kessel, and M.~Lindroos, ``Ai holds promise to drive standards of care, cut healthcare waste,'' \url{https://www.adamsstreetpartners.com/insights/ai-holds-promise-to-drive-standards-of-care/}, January 2024, accessed: 2025-04-09.

\bibitem{who2020}
{World Health Organization}, ``Global strategy on digital health 2020--2025,'' \url{https://www.who.int/docs/default-source/documents/gs4dhdaa2a9f352b0445bafbc79ca799dce4d.pdf}, 2020, accessed: 2025-04-09.

\end{thebibliography}
%





\end{document}